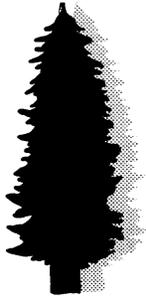

# Cold + Hot Dark Matter Cosmology
# with $m(\nu_\mu) \approx m(\nu_\tau) \approx 2.4$ eV

Joel R. Primack
*Santa Cruz Institute for Particle Physics*
*University of California, Santa Cruz, CA 95064*

Jon Holtzman
*Lowell Observatory, Mars Hill Road, Flagstaff, AZ 86100*

Anatoly Klypin
*Department of Astronomy, New Mexico State University*
*Las Cruces, NM 88001*

David O. Caldwell
*Department of Physics, University of California*
*Santa Barbara, CA 93106-9530*

## Abstract

Cold + Hot Dark Matter (CHDM) $\Omega = 1$ cosmological models require a total neutrino mass $\sim 5$ eV. Because recent data support the $\nu_\mu \to \nu_\tau$ oscillation explanation of the cosmic ray $\nu_\mu$ deficit, which requires that $m(\nu_\mu) \approx m(\nu_\tau)$, this suggests that $m(\nu_\mu) \approx m(\nu_\tau) \approx 2.4$ eV. The linear calculations and N-body simulation reported here indicate that an $\Omega = 1$ CHDM model with two 2.4 eV neutrinos (designated C$\nu^2$DM) agrees remarkably well with all available observations, but only if the Hubble parameter $h \approx 0.5$. We also show that even one 2.4 eV neutrino raises serious difficulties for low-$\Omega$ flat CDM models.

14.60.Pq, 95.30.Cq, 98.80Cq, 98.65.Dx

Predictions of a Cold + Hot Dark Matter (CHDM) cosmological model with a single massive neutrino and $\Omega_\nu = 0.3$ (corresponding to $m_\nu = 94h^2\Omega_\nu = 7$ eV for Hubble parameter $h \equiv H_0/(100 \text{ km s}^{-1} \text{ Mpc}^{-1}) = 0.5$) have been shown [1–3] to agree well with observations, with the possible exception that galaxies may form too late to account for the observations of quasars and damped Ly$\alpha$ systems [4] at high redshifts $z \gtrsim 3$ [5,6]. The latter observations can be accommodated [7] if the assumed $\nu$ mass in CHDM is lowered from $\sim 7$ eV to $\sim 5$ eV. Lowering the $\nu$ mass in CHDM may also give a better account of the Void Probability Function [8] and of the properties of galaxy groups [3,9]; but with one $\sim 5$ eV $\nu$ CHDM probably overproduces clusters, as we show below.

Current experimental data suggest that the net $\nu$ mass is shared among two species of neutrinos. Here we consider the consequences of such $\nu$ masses for the formation of galaxies and large scale structure for cosmological models which are spatially flat and in which most of the DM is cold. We show that $\Omega_\nu = 0.2$ CHDM with the mass evenly shared between two $\nu$ species — C$\nu^2$DM — agrees better with observations than the one-neutrino version, better indeed than any other variant of CDM that we have considered. We first summarize the implications of the latest experimental hints regarding $\nu$ masses, and then compare our calculations of cosmological predictions of various models with observations.

It has been pointed out [10,11] that if the solar $\nu_e$ and atmospheric $\nu_\mu$ deficits arise from the existence of $\nu$ masses, there are only two viable patterns for those masses: (A) the three active neutrinos are approximately degenerate; or (B) the nearly degenerate $\nu_\mu$ and $\nu_\tau$ constitute the hot dark matter, and the $\nu_e$ and a sterile neutrino $\nu_s$ are lighter and are also nearly degenerate [12]. If one also takes into account the need for about 5 eV of neutrino mass for CHDM cosmological models to be viable, pattern (A) corresponds to $m_{\nu_e} \approx m_{\nu_\mu} \approx m_{\nu_\tau} \approx 1.6$ eV, while pattern (B) requires $m_{\nu_\mu} \approx m_{\nu_\tau} \approx 2.4$ eV.

A $\nu$ mass explanation of the solar $\nu_e$ deficit, which is now fairly convincing, implies $\nu_e \to \nu_\mu$ or $\nu_e \to \nu_s$ with $\Delta m_{ei}^2 \equiv |m(\nu_e)^2 - m(\nu_i)^2| \approx 10^{-5}$ eV$^2$ between either pair of particles. Similarly, evidence for a $\nu$ mass explanation of the deficit of $\nu_\mu$'s relative to $\nu_e$'s in atmospheric secondary cosmic rays is also increasing, with compatible results from three experiments [13], and especially new information from Kamiokande [14]. The latter includes accelerator confirmation of the ability to separate $\nu_e$ and $\nu_\mu$ events, as well as an independent higher energy data set giving not only a $\nu_\mu/\nu_e$ ratio agreeing with the lower energy data, but also a zenith-angle (hence source-to-detector) dependence compatible with $\nu_\mu \to \nu_e$ or $\nu_\mu \to \nu_\tau$ oscillations with $\Delta m_{\mu i}^2 \approx 10^{-2}$ eV$^2$. However, almost the entire region of $\Delta m_{\mu e}^2 - \sin^2 2\theta_{\mu e}$ allowed by the Kamiokande data is excluded by data from the Bugey and Krasnoyarsk reactor $\nu$ oscillation experiments. Moreover, the absolute calculated $\nu_e$ and $\nu_\mu$ fluxes — backed by measurements of $\mu$ fluxes — agree with $\nu_e$ data but show a $\nu_\mu$ deficit [15]. Thus $\nu_\mu \to \nu_\tau$ oscillations are favored as an explanation of the atmospheric $\nu_\mu$ deficit.

That the $\nu$ mass pattern (B) might be correct is indicated by early results of the LSND experiment [16], the initial run of which showed an excess of about 8 beam-on events of a type which could be interpreted as $\bar\nu_\mu \to \bar\nu_e$, whereas a background of $\lesssim 1$ event mimicking a $\bar\nu_e$ was expected. The LSND collaboration at this time is not claiming to have observed $\nu$ oscillations, preferring to await results of their current run, which should give three times as much data. Nevertheless, the LSND positron energy distribution (if assumed to be from $\bar\nu_\mu \to \bar\nu_e$, $\bar\nu_e + p \to e^+ + n$) appears [17] to be compatible with scheme (B) and not scheme (A), since the mass-squared difference required is $\Delta m^2_{\mu e} \sim 6$ eV$^2$. This particular value is not in conflict with the KARMEN [18] or BNL E776 [19] experiments, which are least sensitive at that $\Delta m^2$ where LSND is most sensitive. If the $\nu_e$ mass is relatively small ($\lesssim 1$ eV, as indicated for Majorana $\nu$ mass from neutrinoless double beta decay experiments), then the $\nu_\mu$ mass is $\sim 2.4$ eV. This and the $\nu_\mu \to \nu_\tau$ explanation of the atmospheric $\nu_\mu$ deficit then makes $m(\nu_\mu) \approx m(\nu_\tau) \approx 2.4$ eV. It is this scenario for the hot dark matter in a CHDM cosmology which we will show below gives predictions that are in remarkable agreement with astronomical observations.

COBE observations [20] of fluctuations in the microwave background radiation provide an upper limit (since they include possible tensor gravity wave as well as scalar density wave contributions) on the normalization of the spectrum of fluctuations. When this normalization is used for the standard Cold Dark Matter (CDM) model [21] in a critical density ($\Omega = 1$) universe with a Zel'dovich primordial power spectrum ($P(k) = Ak^{n_p}$ with $n_p = 1$) as predicted by simple inflationary models, this fits large-scale data but produces too much structure on smaller scales; for example, there are far too many clusters of galaxies [22] and small-scale velocities of galaxies are too large [3,9].

CDM is attractive because of its simplicity and the existence of well-motivated particle candidates (lightest superpartner particle and axion [23]) for the cold DM; moreover, CDM came remarkably close to predicting the COBE signal. So several variations have been tried to patch up the CDM model. Lowering the normalization (introducing "bias") or "tilting" the primordial spectrum (assuming $n_p \approx 0.7$) improves agreement with data on small scales at the cost of poorer agreement on large scales. The variants of CDM [24] that agree best with observations add either a cosmological constant ($\Lambda$CDM) or a little hot (neutrino) dark matter (CHDM).

We report (quasi-)linear estimates for various observable quantities in Table I. All models in the Table are normalized to COBE except for the CDM model labelled "biased". Actually, our COBE normalization ($Q = 17\,\mu$K) is about 10% lower than the latest analyses [20] would suggest. We have chosen this normalization to allow for a little gravity wave contribution and tilt (consistent with the expectations from simple models of cosmic inflation), and we regard it as being both more realistic and more conservative than a higher normalization — more conservative, since the greatest problem for the CHDM models is enough early structure formation, a problem that worsens as the normalization of the spectrum is decreased.

The first two lines of numbers give our estimates of a variety of observational quantities and the uncertainties in them, from large to small scales. The bulk velocity at $r = 50\,h^{-1}$ Mpc is derived from the latest POTENT analysis [25]; the uncertainty includes the error from the analysis but not cosmic variance. However, similar constraints come from other data on large scales such as power spectra that may be less affected by cosmic variance since they probe a larger volume of the universe. We have estimated the current number density of clusters ($N_{\rm clust}$) from comparison of data on the cluster temperature function from X-ray observations with hydrodynamic simulations [26] as well as from number counts of clusters [27]. All recent estimates of the cluster correlation function give fairly large values at $30\,h^{-1}$ Mpc [28]; this also suggests that the zero crossing of the correlation function must exceed $\sim 40\,h^{-1}$ Mpc. The linear estimate of pairwise velocities ($\sigma_v$) is not an observed value, since pairwise velocities are strongly influenced by nonlinear evolution. However, from experience with N-body simulations for various models, we have found that the results from simulations are roughly a factor of 3 larger than the linear estimate. We have confirmed that factor by such a simulation for the $C\nu^2 DM$ model (high resolution $800^3$ PM mesh in a 50 $h^{-1}$ Mpc box with $256^3$ cold and $2 \times 256^3$ hot particles), but all values given here are for linear calculations. The limit we give is therefore our estimate of the maximum linear value allowed by observations. The final column gives the observed density in cold hydrogen and helium gas at $z = 3.0 - 3.5$ from observations of damped Lyman $\alpha$ systems [4].

The next two lines present predictions from the CDM model, and illustrate its problems. The cluster correlation function at 30 $h^{-1}$ Mpc is smaller than observations indicate regardless of CDM normalization, reflecting the fact that the matter correlation function becomes negative beyond 36 $h^{-1}$ Mpc. If CDM is normalized to $\sigma_8 = 0.7$ (or equivalently to linear bias $b \equiv \sigma_8^{-1} = 1.43$), the cluster density problem is avoided, but small-scale velocities are still too large and bulk velocities on a scale of 50 $h^{-1}$ Mpc are probably too low.

The problem with CDM is that it has too much power on small scales relative to power at large scales. Since including hot DM reduces small scale power (because $\nu$ free streaming causes perturbations to damp on smaller scales), including a $\nu$ component improves the agreement with observations. The model studied by KHPR [1] with a $\nu$ mass of 7 eV, corresponding to $\Omega_\nu = 0.3$ for $h = 0.5$, is a much better match to observations than CDM, but it has $\Omega_{\rm gas}$ too small [6,7]. The small-scale velocities in this model are small enough [1] to agree with the old result $\sigma_v(1\ h^{-1}$ Mpc$) = 340$ km s$^{-1}$ from the CfA1 survey [29]. However, this result is now known to be in error because of the accidental omission of the Virgo cluster [30]; moreover, the $\sigma_v$ statistic is not very robust [31], since it is heavily influenced by the presence of (relatively rare) clusters. A direct comparison of galaxy groups in "observed" CDM and CHDM simulations with identically selected CfA1 groups shows that CDM velocities are much too high, even with biasing, while the velocities in the $\Omega_\nu = 0.3$ CHDM model are in reasonable agreement [3,9]. However, the fraction of galaxies in groups is slightly too high for $\Omega_\nu = 0.3$ CHDM, while it is significantly too low for CDM. Thus agreement is improved for a lower $\Omega_\nu$.

Lowering $\Omega_\nu$ to 0.20 (1$\nu$) increases small-scale velocities but not enough to conflict with the data, and it raises $\Omega_{\rm gas}$ enough for early object formation [7]. However, this model probably overproduces clusters. In order to avoid this, it could be normalized lower, given some gravity wave contribution to COBE, but this would result in too little early structure formation.

Now consider $C\nu^2 DM$. All quantities are in good agreement with the astronomical data if the same $\Omega_\nu = 0.2$ is divided between two $\nu$ species, as suggested by the data. The ratio of the power spectrum for $C\nu^2 DM$ compared to that for CHDM with the same total $\nu$ mass in one species is $\sim 1$ at large and small scales, but it has a dip of about 30% centered at $\sim 10\ h^{-1}$ Mpc. The larger $\nu$ free-streaming length, resulting from a $\nu$ mass of 2.4 eV instead of twice that, lowers the abundance of clusters and agrees better with observations.

It is remarkable that, with the experimentally suggested $\nu$ masses, only cosmological models with $h \approx 0.5$ match observations. Returning to Table I, note that for $h = 0.7$ — favored by many observers — CDM (CDM$_{0.7}$) is an even worse fit to the data than for $h = 0.5$ because the larger $h$ makes matter-dominance ($\propto \Omega h^2$) occur earlier and thus moves the bend in the CDM spectrum to smaller scales, giving more intermediate and small scale power for a given large scale normalization. Adding two 2.4 eV neutrinos (C$\nu^2$DM$_{0.7}$) only slightly improves the situation, because this only gives $\Omega_\nu \sim 0.1$ for $h = 0.7$, so the spectrum is not modified very much. (Recall that for given $m(\nu)$, $\Omega_\nu$ scales as $h^{-2}$ since critical density is $\propto h^2$.) The $h = 0.7$ model can match data better with a tilted spectrum of primordial fluctuations with $n_p = 0.81$ (C$\nu^2$DM$_{n0.8}$) — but only without gravity waves. Typical cosmic inflation models with this much tilt would suppress scalar power by a factor of $\sim 0.5$ leading to very serious underproduction of clusters and of $\Omega_{\text{gas}}$, although models have been proposed [32] with tilt but no gravity waves. Of course, with large $h$, $\Omega = 1$ leads to too short a time since the big bang: $t_0 = \frac{2}{3}H_0^{-1} = 6.52\,\text{Gy}/h = 9.3$ Gy for $h = 0.7$.

A larger age is obtained for $\Omega_0 < 1$ which, to be consistent with inflation, requires a positive cosmological constant $\Lambda$. The maximum value allowed by the COBE data is $\Omega_\Lambda \equiv \Lambda/(3H_0^2) \approx 0.78$ [33], and the maximum allowed by quasar lensing statistics is $\Omega_\Lambda \approx 0.7$ [34]. For a flat ($k = 0$) universe with $\Omega_\Lambda = 0.7$ and $\Omega_0 = 0.3$, $h = 0.7$ corresponds to $t_0 = 13.5$ Gy. $\Lambda$CDM with these parameters is a fairly good fit [35] to the data, although our linear calculations suggest that not enough clusters are produced and bulk velocities may be too low. However, this model becomes much worse if even one $\nu$ of 2.4 eV is added, seriously underproducing clusters and $\Omega_{\text{gas}}$.

A similar situation occurs for $\Omega = 1$ C$\nu^2$DM with $h = 0.4$, for which $\Omega_\nu = 0.32$ with two 2.4 eV neutrinos. Because the bend in the CDM spectrum moves to larger scales as $h$ decreases, there is less small scale power for given large scale normalization; adding hot DM further decreases small scale power. We find that even with only one 2.4 eV $\nu$, there is just not enough power to generate the observed number of clusters or high-redshift objects.

Ever since the early 1980s there have been hints [36] that features on small and large scales may require a hybrid scenario in which there are two different kinds of dark matter. Preliminary studies of the CHDM scenario were carried out in 1984 [37], and it was first worked out in detail only in the last two years [1–3] with one massive $\nu$. We have shown here that the C$\nu^2$DM model, with Hubble parameter $h = 0.5$ and both neutrinos having a mass of 2.4 eV as suggested by ongoing experiments, gives a remarkably good account of all presently available astronomical data. New data on CMB, large scale structure, and structure formation will severely test this highly predictive model. Experimental results expected soon will clarify whether indeed $m(\nu_\mu) \approx m(\nu_\tau) \approx 2.4$ eV. Table I shows the implications of such $\nu$ masses for a variety of popular CDM-type cosmological models. If even just the $\nu_\mu$ has a mass of 2.4 eV, as suggested by preliminary results from the LSND experiment, flat low-$\Omega$ CDM models are disfavored.

ACKNOWLEDGEMENTS. We have benefited from conversations with S. Bonometto, S. Borgani, A. Dekel, S. Faber, R. Nolthenius, and D. Seckel. This research was supported by NSF grants at UCSC and NMSU and by a DOE grant at UCSB. The simulation was done on the Convex C-3880 at NCSA.

TABLES

TABLE I. Comparison of models: COBE normalization ($Q_{\rm rms-ps} = 17\mu K$).

| Model | $\Omega_{\rm bar}$ (%) | $\Omega_\nu$ (%) | $N_\nu$[a] | $m_\nu$[a] | $\sigma_8$[b] | $V$[c] 50Mpc | $N_{\rm clust}$[d] ($10^{-7}$) | $\xi_{cc}$[e] 30Mpc | $r$[f] $\xi = 0$ | $\sigma_v$[g] 1Mpc | $\Omega_{\rm gas}$[h] ($10^{-3}$) |
|---|---|---|---|---|---|---|---|---|---|---|---|
| OBSERVATIONS | | | | | | 335 | 4.0 | 0.30 | > 40 | < 200 | 6.0 |
| uncertainties | | | | | | 80 | 2.0 | 0.15 | | | 2.0 |
| CDM models, h=0.5 | | | | | | | | | | | |
| COBE | 7.5 | 0 | 0 | 0.00 | 1.08 | 356 | 48 | 0.10 | 36 | 405 | 30 |
| biased | 7.5 | 0 | 0 | 0.00 | 0.70 | 231 | 1.2 | 0.08 | 36 | 262 | 14 |
| CHDM models, h=0.5 | | | | | | | | | | | |
| C$\nu^2$DM | 7.5 | 20 | 2 | 2.35 | 0.67 | 347 | 2.4 | 0.35 | 70 | 144 | 4.6 |
| KHPR | 10.0 | 30 | 1 | 7.04 | 0.66 | 359 | 3.6 | 0.37 | 51 | 98 | 0.4 |
| 1$\nu$ | 7.5 | 20 | 1 | 4.69 | 0.75 | 357 | 7.7 | 0.30 | 52 | 156 | 5.2 |
| CDM/CHDM models, h=0.7 | | | | | | | | | | | |
| CDM$_{0.7}$ | 5.0 | 0 | 0 | 0.00 | 1.56 | 393 | 180 | -0.01 | 28 | 714 | 31 |
| C$\nu^2$DM$_{0.7}$ | 4.0 | 10 | 2 | 2.30 | 1.24 | 389 | 93 | 0.09 | 38 | 432 | 20 |
| C$\nu^2$DM$_{n0.8}$ | 4.0 | 10 | 2 | 2.30 | 0.71 | 271 | 2.1 | 0.14 | 49 | 199 | 4.4 |
| $\Lambda$CDM/$\Lambda$CHDM models, $h = 0.7$, $\Omega_0 = 0.3$, and $\Omega_\Lambda = 0.7$ | | | | | | | | | | | |
| $\Lambda$CDM | 2.6 | 0 | 0 | 0.00 | 0.86 | 277 | 0.23 | 0.20 | 125 | 113 | 12 |
| $\Lambda$CHDM | 2.6 | 5 | 1 | 2.30 | 0.54 | 263 | 5E-4 | 0.35 | 136 | 48 | 0.4 |
| $\Lambda$C$\nu^2$DM | 2.6 | 10 | 2 | 2.30 | 0.33 | 247 | 2E-9 | 0.53 | 144 | 19 | 6E-7 |

[a]$N_\nu$ is the number of $\nu$ species with mass. If $N_\nu \geq 1$, each species has the same mass $m_\nu$.
[b]$(\Delta M/M)_{\rm rms}$ for $R_{\rm top-hat} = 8h^{-1}$ Mpc.
[c]Bulk velocity in top-hat sphere of radius $50h^{-1}$ Mpc.
[d]Number density of clusters $N(> M)$ in units of $10^{-7} h^3 {\rm Mpc}^{-3}$ above the mass $M = 10^{15}h^{-1}M_\odot$, calculated using Press-Schechter approximation with gaussian filter and $\delta_c = 1.50$.
[e]The cluster-cluster correlation function amplitude at $30h^{-1}$ Mpc, computed using linear theory [24] and assuming a unit bias factor for the dynamical contribution.
[f]Zero crossing ($\xi(r) = 0$) of the correlation function in units of $h^{-1}$ Mpc.
[g]Linear estimate of pairwise velocity at $r = 1h^{-1}$ Mpc scale: $\sigma_v^2 = 2H_0^2 \int dk\, P(k)(1 - \sin kr)/kr$.
[h]Mean density of collapsed baryons at $z = 3-3.5$ in units of $10^{-3}$ of critical density, calculated using $\Omega_{\rm gas} = (\Omega_b/\Omega_c)\, {\rm erfc}(\delta_c/\sqrt{2}\sigma)$, with $\delta_c = 1.4$ [7], and $\sigma$ computed for mass $5 \times 10^{10}h^{-1}M_\odot$ using gaussian smoothing and assuming all gas is neutral. Since some gas may be ionized or removed by star formation, $\Omega_{\rm gas}$ for the various models should be at least as high as the observations.